\documentclass[aps,prb,twocolumn,showpacs,superscriptaddress,groupedaddress]{revtex4}  
\usepackage{graphicx}  
\usepackage{dcolumn}   
\usepackage{bm}        
\usepackage{amssymb}   
\usepackage{amsmath}

\begin{document}


\title{Whether it is possible to stabilize the 1144-phase pnictides with tri-valence cations?}
\author{B. Q. Song}
\affiliation{Ames Laboratory, US Department of Energy and Department of Physics and Astronomy, Iowa State University, Ames, Iowa, 50011, USA}    
\author{Manh Cuong Nguyen}
\affiliation{Ames Laboratory, US Department of Energy and Department of Physics and Astronomy, Iowa State University, Ames, Iowa, 50011, USA}    
\author{C. Z. Wang}
\affiliation{Ames Laboratory, US Department of Energy and Department of Physics and Astronomy, Iowa State University, Ames, Iowa, 50011, USA}    
\author{P. C. Canfield}
\affiliation{Ames Laboratory, US Department of Energy and Department of Physics and Astronomy, Iowa State University, Ames, Iowa, 50011, USA}
\author{K. M. Ho}
\affiliation{Ames Laboratory, US Department of Energy and Department of Physics and Astronomy, Iowa State University, Ames, Iowa, 50011, USA}  

\date{\today}

\begin{abstract}
The 1144 iron arsenide (e.g. $CaKFe_{4}As_{4}$) has recently been discovered and inspired a tide of search for superconductors. Such far, the discovered compounds are confined to iron arsenides ($ABFe_{4}As_{4}$), where $A$ and $B$ are either alkali metals or alkaline earth elements. In this work, we propose two directions in searching 1144 structures: (i) using tri-valence cations for $A$; (ii) substituting the transition metal, e.g. replacing Fe by Co. Following the two directions, we employ density functional theory to study stability and electronic structures of 1144 pnictides of various tri-valence cations (La, Y, In, Tl, Sm and Gd), as well as cobalt arsenides. For $LaAFe_{4}As_{4}$, the 1144 phase can be stabilized in three systems: $LaKFe_{4}As_{4}$, $LaRbFe_{4}As_{4}$ and $LaCsFe_{4}As_{4}$, which show quasi-two-dimensional semi-metal features similar to the iron pnictide superconductors: hole-type Fermi surface at $\Gamma$ point and electron-type Fermi surfaces at M point in B.Z. In addition, $LaKFe_{4}As_{4}$ feature an extra “bubble shaped” Fermi surface sheets, distinct from the other two peers. Y does not support any 1144 phase within our search. For In and Tl, substitute Fe by Co and two unknown compounds of the 122 phase are stabilized: $InCo_{2}As_{2}$ and $TlCo_{2}As_{2}$. The two cobalt arsenides have Fermi surfaces of similar topology as iron arsenides, but the Fermi surfaces are all electron-type, showing potentials to be undiscovered superconductors. Stable 1144 phases are also found in $InKCo_{4}As_{4}$ and $InRbCo_{4}As_{4}$. For Sm and Gd, most 1144 or 122 iron arsenides are found unstable. 
\end{abstract}

\maketitle

\section{Introduction}
Since $LaFeAsO_{1-x}F_{x}$ was discovered [1], subsequent works lead to synthesis of various iron pnictides superconductors(SC), belonging to several structural families: 1111, 111, 122, etc. [2,3] These crystalline structures commonly feature Fe-pnictogen skeleton layers, where each Fe is coordinated with four pnictogen atoms in a tetrahedral way. The common feature provides a structural prototype in designing SC. The newly discovered 1144-phase iron arsenide ($ABFe_{4}As_{4}$) [6-8] belongs to such structural prototype (Figure 1a) and shows superconductivity at temperatures comparable to the 122 phase [6-8]. This inspires great interest in searching SC in the 1144 phase [6-8,10-19].

\begin{figure} [t]
\includegraphics[scale=0.6]{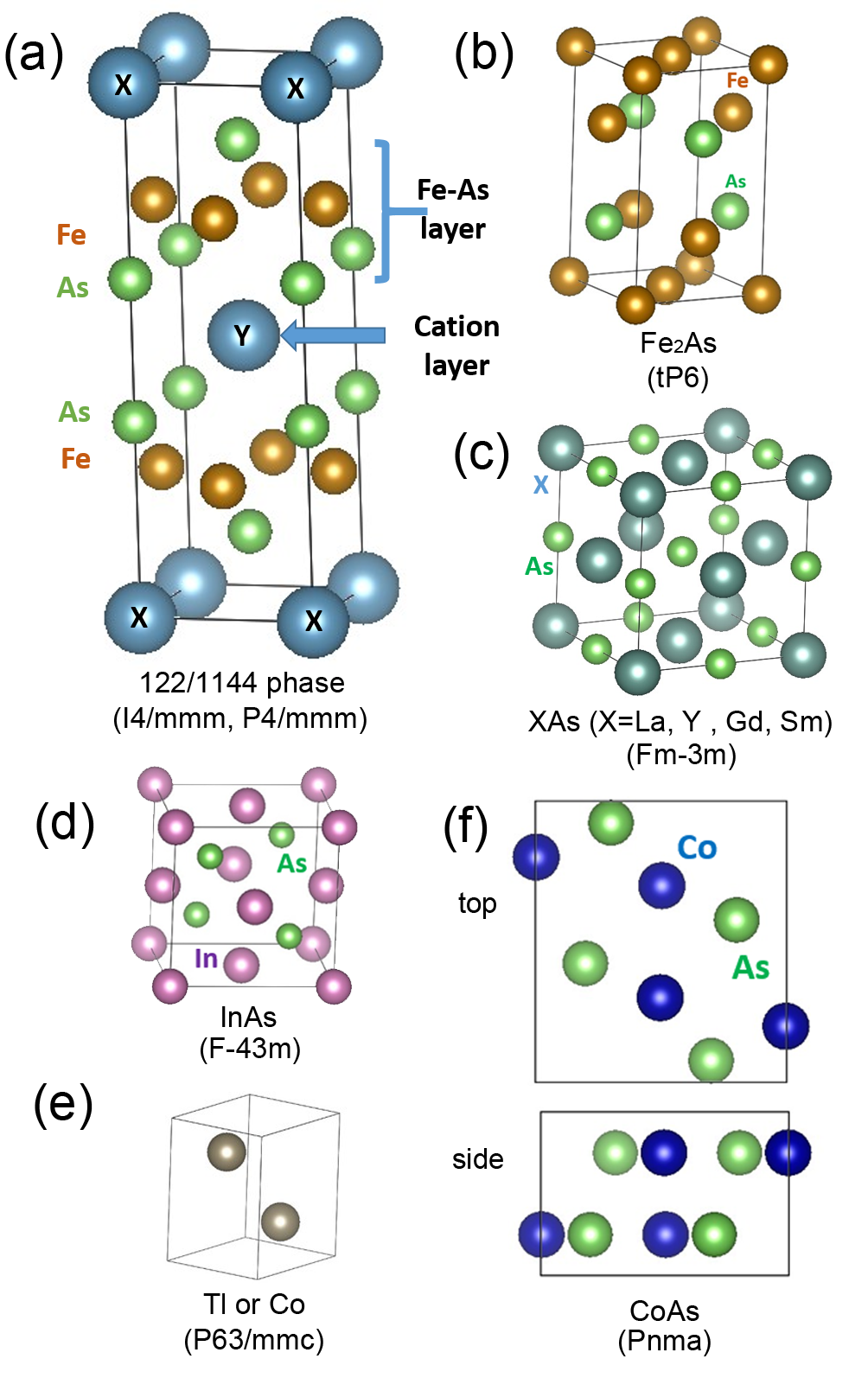}
\caption{\label{fig:epsart} (Color online) Crystal structures involved. (a) 122/1144 phase. 122 phase if $X$ $Y$ sites are occupied by same cations. 1144 phase if $X$ and $Y$ are occupied by different cations. It consists of two subsystems: Fe-As layer and cation layer. (b) $Fe_{2}As$. (c) $XAs$. (d) $InAs$. (e) $\alpha$-phase of Tl or Co. (f) The top and side views of $CoAs$ unit cell. The space group is labelled on each graph.}
\end{figure}

The 1144 phase $ABFe_{4}As_{4}$ (Figure 1) has the same chemical composition as the 50\%-doped 122 phase ($A_{0.5}B_{0.5}Fe_{2}As_{2}$), but distinct in the cation layers. For $A_{0.5}B_{0.5}Fe_{2}As_{2}$, the cations $A$ and $B$ are randomly distributed and form a solution phase, which has the same space group as the pure 122 phase. Such solution phase is referred to as 122(s) phase in this context. While the 1144 phase is obtained when cation layers are alternatively occupied by $A$ and $B$. Thus, the 1144 phase is structurally ordered and having a lower symmetry compared with the 122(s) phase. In experiment, both the two phases have been observed [6], and which one is more stable depends on specific choice of $A$-$B$ combinations. The stability of a particular $A$-$B$ combination in the 1144 iron arsenides can be characterized by two parameters: the mismatch in lattice constants ${\Delta}a$ and the mismatch in atomic size ${\Delta}r$ [6] (or ${\Delta}r$ could be replaced with lattice mismatch ${\Delta}c$ [27]). The 1144 phase tends to be stabilized with a large ${\vert}{\Delta}r{\vert}$ and small ${\vert}{\Delta}a{\vert}$. Nevertheless, discovered compounds are currently confined within cations of alkali metals (IA group), alkaline earth (IIA group) elements [6-8,20] and Eu [21-22].

In this work, we propose two directions in searching new 1144 compounds: (i) build tri-valence cations into the 1144 phase; (ii) Substitute the transition metal, e.g. Fe replaced by Co. For the first direction, we examine six tri-valence elements: Y La In Tl Sm and Gd, which can be classified into three groups. The first group includes Y ($4d^{1}5s^{2}$) and La ($5d^{1}6s^{2}$). They are non-magnetic and similar with IA or IIA elements in the sense of atomic size and a low electronegativity about 1.0. The second group are In and Tl. Despite the similar electronic shell structure as IA and IIA, the electronegativity (e.g. In 1.78, Tl 1.62 based on Pauling scale) is apparently higher than IA or IIA atoms (generally below 1.0). A higher electronegativity allows In (or Tl) to compete with Fe in forming chemical bonding with As. Consequently, a stable binary alloy InAs usually forms, impeding the formation of Fe-As layers. The third group are Sm and Gd. They are the neigborhood of Eu in periodic table (Eu can form stable 122 iron arsenide) and similar with Eu in terms of shell structure, electronegativity and atomic size. For the second direction, we notice that even most tri-valence elements, for instance, La does not have stable 122 iron arsenide phase $LaFe_{2}As_{2}$, but the 122 phase will be stabilized if Fe is replaced by Co or Ru. We suggest that this strategy might be applied to other tri-valence elements. 

In Sec.II we describe our methodology and calculation details. In each following section, one category of tri-valence elements are discussed. In Sec.III, we discuss building La and Y into 1144 structures, specifically focusing on $LaAFe_{4}As_{4}$ ($A$=Na, K, Rb, Cs, Ca, Sr, Ba). In Sec.IV, In and Tl are discussed. We propose stabilizing the 1144 structure by substituting Fe by Co. Thus we investigate $TlCo_{2}As_{2}$ and $InCo_{2}As_{2}$. We also study $TlACo_{4}As_{4}$ and $InACo_{4}As_{4}$ ($A$=K, Rb, Ca, Sr, Ba). In Sec.V, we will discuss Sm and Gd in the 1144 phase. 

\section{Methodology and calculation details}

At zero temperature, the solution phase does not exist for the vanishing entropy. Hence, to study the phase stability, one has only to examine formation enthalpy of the 1144 phase with respect to phase decomposition, which varies with systems, as shown in phase diagrams (Figure 2).

\begin{figure}
\includegraphics[scale=0.8]{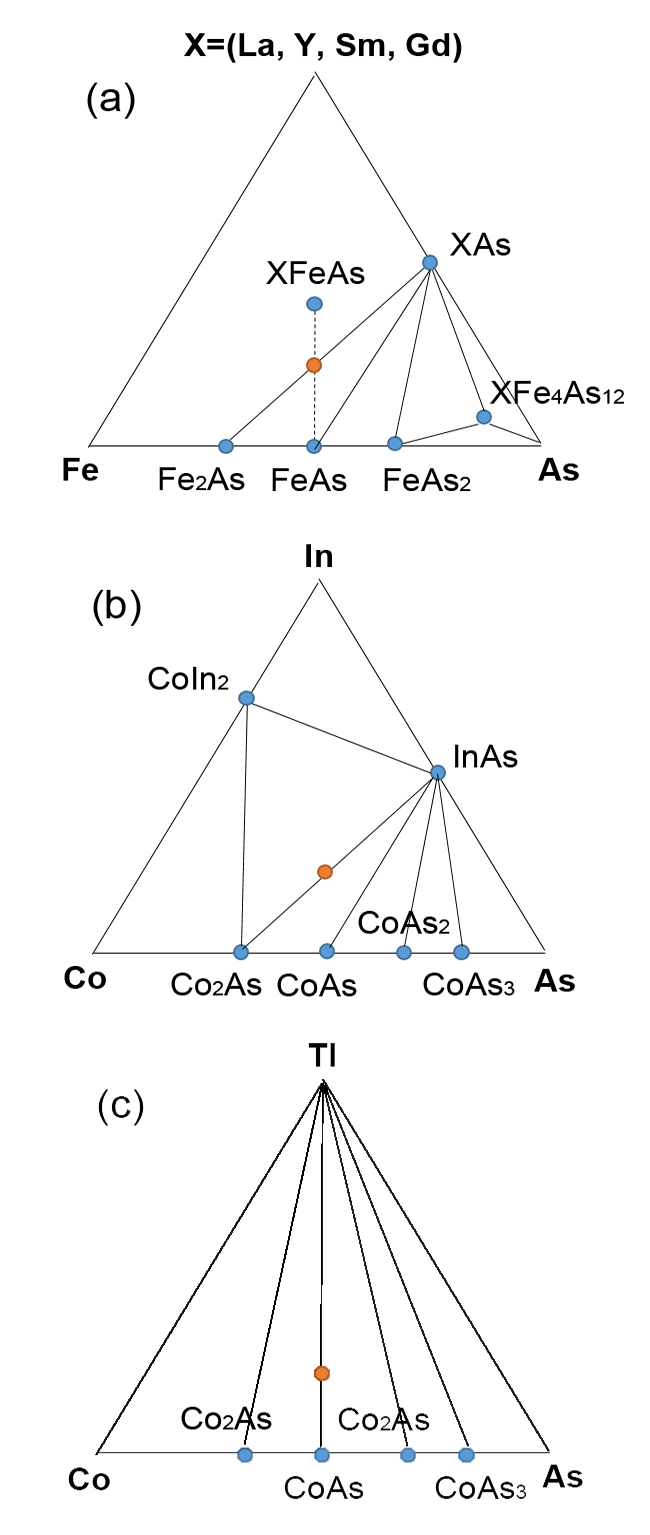}
\caption{\label{fig:epsart} (Color online) The phase diagram of several ternary alloys: (a) X-Fe-As (X=La, Y, Sm and Gd); (b) In-Co-As; (c) Tl-Co-As. The known thermodynamically stable phase is denoted by blue dots on the graph. The 122 phase is designated as the red dot.}
\end{figure}

At finite temperature, the solution phase comes into play, which complicates the problem in two ways. First, in estimating energy, one needs to consider all configurations in the solution phase [6,23-26]. In that case, supercell modelling is apparently unfeasible. Thus, we adopt an approximation of ideal solution:
\begin{equation}
E_{sol}=x{\cdot}E_{AFe2As2}+(1-x){\cdot}E_{BFe2As2}
\end{equation}
where $E_{sol}$ is the energy of solution $A_{x}B_{1-x}Fe_{2}As_{2}$ and $x$ is the concentration of $A$ cations. Second, one is obliged to consider the entropy contribution to the free energy. There are several sources for entropy: configurational, vibrational and electronic. Earlier studies showed that the configurational entropy plays the most important role [27]. Hence, we only consider configurational entropy and adopt an ideal solution approximation to estimate it:[39]

\begin{equation}
S=-2k_{B}(x{\log}x+(1-x){\log}(1-x))
\end{equation}

With these approxations, one is enabled to estimate the critial transition temperature for 1144-122(s) phases and La-solubility at finite temperature. These results can be compared with experiment observations. For example, the mixture of La-Ca will lead to a La-doped $CaFe_{2}As_{2}$ up to 20\% [28]. We will detail our calculation in Sec. III B.

All calculations of energy are based on density functional theory (DFT) and Perdew Burke Ernzerhof (PBE) correlation functional [29], implemented by the Vienna ab initio simulation package (VASP) [30]. The projected augmented wave pseudopotential method [31] is employed. The k-sampling has also been tested to ensure a stable energy. In calculating the enthalpy difference, we adopt spin-polarized calculation to account the spin configurations. In calculating density of states (DOS) and band structures, we adopt a non-spin-polarized calculation, which has been commonly used in previous studies [32,33]. 

A single 1144 unit cell contains 10 atoms: one $A$, one $B$, four Fe and four As atoms (Figure 1a). Both experiments and calculations show that 122 and 1144 structures have ground states of stripe anti-ferromagnetic (AFM) coupling [34] within Fe-As layer and simple AFM coupling between Fe-As layers. Hence, an enlarged 2$\times$2$\times$1 supercell of $ABFe_{4}As_{4}$ are constructed to account for the magnetism. In Sec. IV we will consider cobalt-arsenide, whose magnetism is less well known and more complicated. For example, $SrCo_{2}As_{2}$ [35] and $BaCo_{2}As_{2}$ [36] do not show any magnetic ordering down to 2K. That probably indicates AFM and FM should be degenerate in energy. While, $CaCo_{2}As_{2}$ undergoes an AFM magnetic transition at 76K and magnetic moment is aligned in z-axis. [37] Magnetism of $KCo_{2}As_{2}$ and $RbCo_{2}As_{2}$ is not well known. In addition, recent study provides evidence of magnetic fluctuation of stripe-AFM and FM type in $SrCo_{2}As_{2}$, despite absence of magnetic ordering [38]. In our work, we adopt AFM for $CaCo_{2}As_{2}$ and FM for other four cobalt arsenides in calculating structural stability. In calculating band structures, we have adopted a non-magnetic scenario. 

\section{La- and Y-contained 1144 structures.}
In this section, we will explore the stability of La- and Y-contained 1144 structures, as well as their electronic structures. 

\subsection{Phase stability at zero temperature}
The phase diagram for La(Y)-Fe-As is presented in Figure 2a. Note that the 122 phase tends to decompose: $YFe_{2}As_{2}{\rightarrow}YAs + Fe_{2}As$ and $LaFe_{2}As_{2}{\rightarrow}LaAs + Fe_{2}As$. Thus we can define the enthalpy difference as: 
\begin{equation}
\begin{split}
{\Delta}H^{122}&=E_{LaFe2As2}-(E_{LaAs}+E_{Fe2As}) \\
{\Delta}H^{122}&=E_{YFe2As2}-(E_{YAs}+E_{Fe2As})
\end{split}
\end{equation}
which are listed in the first row of Table I. We find positive ${\Delta}H$ for both $YFe_{2}As_{2}$ and $LaFe_{2}As_{2}$, which confirms the information given by the phase diagram. To prevent phase decomposition, we propose using a partner element (e.g. IA or IIA group elements), to form the La- or Y-contained 1144 phase. This idea is inspired by our earlier work [27], which suggests that the favorable valence state of cations in forming the 1144 phase is +1.5. Thus, one needs to effectively lower the valence state of $La^{3+}$. In that sense, IA group elements seem more advantageous.

\begin{table}
\caption{\label{tab:table1}The formation enthalpy of $LaFe_{2}As_{2}$, $YFe_{2}As_{2}$ (the first row in the table) and 1144 iron arsenide $LaAFe_{4}As_{4}$, $YAFe_{4}As_{4}$. The enthalpy of $LaFe_{2}As_{2}$ and $YFe_{2}As_{2}$ are measured with respect to phase decomposition: $Fe_{2}As+MAs$ ($M$=La, Y). The formation enthalpies of La- and Y-contained 1144 structures are defined with respect to phase decomposition: $Fe2As+LaAs+AFe_{2}As_{2}$ ($A$=Ca, Sr, Ba, Na, K, Rb, Cs). Positive ${\Delta}H$ favors phase decomposition. Note that ${\Delta}H$ is defined differently for 122 and 1144 phases.}
\begin{ruledtabular}
\begin{tabular}{c c c c c c}
 & Lattice & ${\Delta}H$ & & Lattice & ${\Delta}H$ \\
 & ($\AA$) & (meV/atom) & & ($\AA$) & (meV/atom) \\
La & a=3.8427 & 19.54 & Y & a=3.7737 & 86.51 \\
				 & c=11.9341 &      & 				  & c=11.5686 &         \\
LaCa & a=3.8427  & 6.73 & YCa & a=3.8195  & 51.13 \\
	 & c=12.0909 &      &     &	c=11.7666 &       \\
LaSr & a=3.8966  & 1.54 & YSr & a=3.8692  & 47.96 \\
	 & c=12.2102 &      &     &	c=11.9188 &       \\
LaBa & a=3.9434  & 0.67 & YBa & a=3.9275  & 41.11 \\
	 & c=12.3938 &      &     &	c=12.0488 &       \\
LaNa & a=3.7974  & -0.39 & YNa & a=3.7370  & 30.92 \\
	 & c=12.5987 &      &     &	c=12.3730 &       \\
LaK  & a=3.8278  & -13.91 & YK & a=3.8125  & 17.81 \\
	 & c=13.1093 &      &     &	c=12.7577 &       \\
LaRb & a=3.8431  & -17.22 & YRb & a=3.8200 & 16.03 \\
	 & c=13.3857 &      &     &	c=13.1058 &       \\
LaCs & a=3.8777  & -18.28 & YCs & a=3.8660 & 12.81 \\
	 & c=13.6276 &      &     &	c=13.2890 &       \\
\end{tabular}
\end{ruledtabular}
\end{table}

To test this idea, we study both IA elements (Na, K, Rb, Cs) as well as IIA elements (Ca, Sr, Ba). The ${\Delta}H$ of 1144 structure is defined as
\begin{equation}
\begin{split}
{\Delta}H^{1144}&=E_{LaXFe4As4}-(E_{XFe2As2}+E_{LaAs}+E_{Fe2As}) \\
{\Delta}H^{1144}&=E_{YXFe2As2}-(E_{XFe2As2}+E_{YAs}+E_{Fe2As})
\end{split}
\end{equation}
Similarly, a positive ${\Delta}H$ will favor phase decomposition. Note that ${\Delta}H$ of the 1144 phase is defined differently from the 122 phase. Also note that there is another way of decomposing: $LaFe_{2}As_{2}{\rightarrow}LaFeAs+FeAs$ (as the dash line shown in Figure 2a). However, we find that this type of decomposition is much higher in energy by 200 meV/atom than the decomposition: $LaFe_{2}As_{2}{\rightarrow}LaAs+Fe_{2}As$. 

\begin{figure}[t]
\includegraphics[scale=0.7]{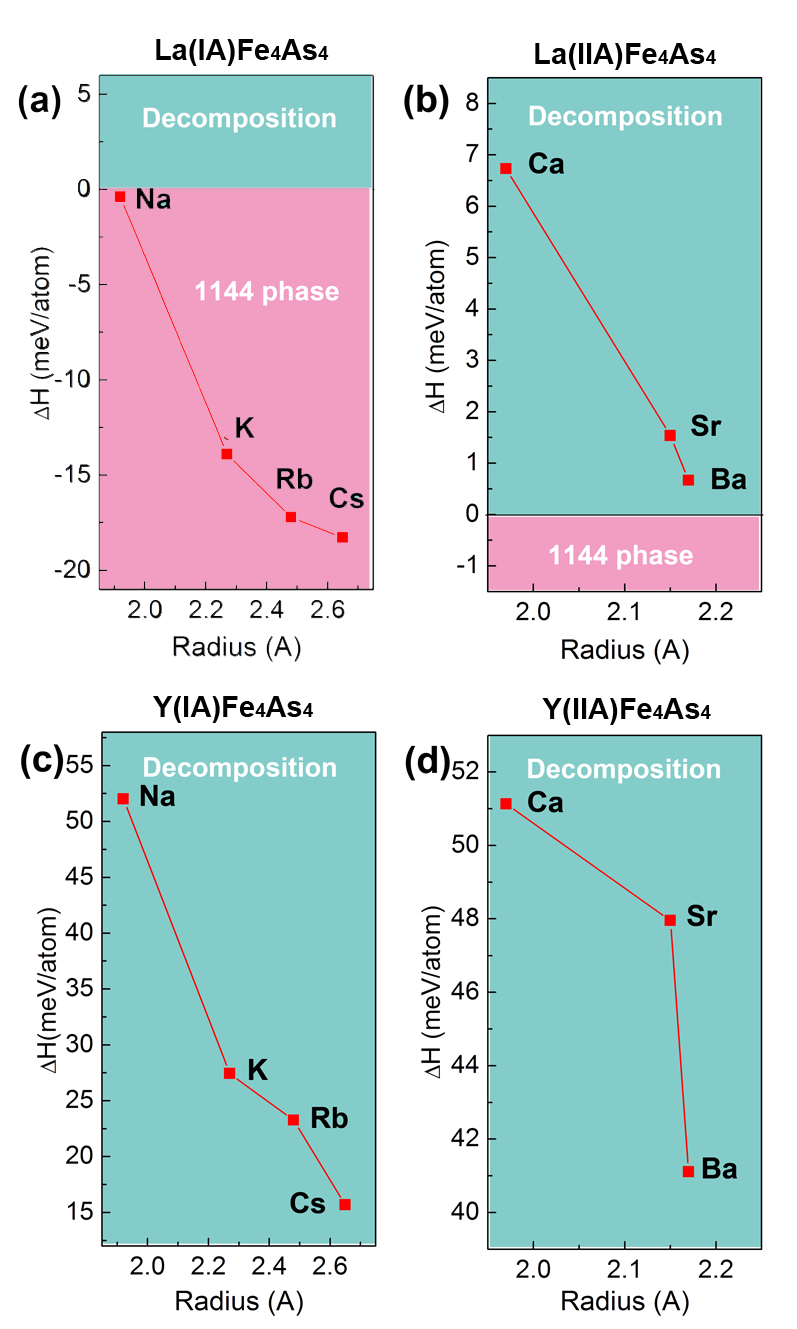}
\caption{\label{fig:epsart} (Color online) The enthalpy dependence on atomic radius of cations in several La- and Y-contained 1144 structures. Postive ${\Delta}H$ favors phase decomposition, colored in blue, while negative ${\Delta}H$ favors 1144 phase, colored in red.}
\end{figure}

The calculated ${\Delta}H$ are listed in Table I. We find four stable 1144 compounds: $LaNaFe_{4}As_{4}$, $LaKFe_{4}As_{4}$, $LaRbFe_{4}As_{4}$, and $LaCsFe_{4}As_{4}$. None of the Y-contained 1144 compounds are found stable. In addition, two trends for ${\Delta}H$ are recognized. First, La+IA combinations better stabilize the 1144 phase than La+IIA, which is consistent with our expectation. Second, the stability of 1144 phase is enhanced with an increased radius of $A$ atom. We plot the ${\Delta}H$ vs atomic radius of $A$ in $LaAFe_{4}As_{4}$ and $YAFe_{4}As_{4}$ in Figure 3. Evidently, ${\Delta}H$ keep decreasing as the radius of cation is increased. Whereas only for La+IA combinations, ${\Delta}H$ arrives into the 1144 phase. Previous work has attributed the stability of the 1144 phase solely to structural mismatches [6], which is well fitted to experimental findings. However, the range of fitted compounds were confined to IA+IIA combinations for cations, which are similar in charge transfer, thus structural factors become outstanding. In fact, charge also plays an important role as shown above. Nevertheless, we find, at zero temperature, four 1144 compounds $LaAFe_{4}As_{4}$ ($A$=Na, K, Rb and Cs) are stable; others will undergo phase decompositions.  

\subsection{Phase stability at finite temperature}
At finite temperature, three matter states compete with each other: the 1144 phase, the phase decomposition ($AFe_{2}As_{2}+LaAs+Fe_{2}As$) and the 122(s) phase. The free energy is $G=E-ST$, where only configurational  entropy is considered. Since the 1144 phase and phases produced by decomposition are entirely ordered, their entropy is zero. On the other hand, 122(s) phase has entropy due to randomness, which is 0.011946 meV/(atom K). Thus at high enough temperature, the 122(s) phase will eventually predominate due to -$ST$. We can estimate the critical temperature for the 122(s) phase overtaking the 1144 phase. We find the following critical temperatures: 45 K for $LaNaFe_{4}As_{4}$, 1615 K for $LaKFe_{4}As_{4}$, 2000 K for $LaRbFe_{4}As_{4}$ and 2123 K for $LaCsFe_{4}As_{4}$. An achievable 1144 phase in experiment requires the critical temperature higher than the room temperature. Because the crystal is grown at temperatures 800 K - 1000 K, then cooling down to room temperature [7]. Thus, the three cation combinations of LaK, LaRb and LaCs seem promising to be synthesized. 

For those unstable combinations ($LaAFe_{4}As_{4}$, $A$=Ca, Sr and Ba), the system will not entirely undergo phase decomposition due to temperature effect. Instead, certain amount of La will dissolve and form a solution phase $(La_{x}A_{1-x})Fe_{2}As_{2}$, owing to the entropy contribution to free energy. Since $x$ would be non-zero values at finite temperature, there is no sharp boundary between the “pure” 122 phase and the 122(s) phase. The pure 122 phase observed in experiment actually should be regarded as a doped 122 phase with extremely low concentration. Then, the question is what the solubility of La is at a given temperature.

To estimate the maximum solubility $x$, we consider a La-rich environment: ($La_{y}A_{1-y}Fe_{2}As_{2}$), ($y>x$). Energy drives the system to decompose into ordered phases: 
\begin{equation}
La_{y}A_{1-y}Fe_{2}As_{2}{\rightarrow}(1-y){\cdot}AFe_{2}As_{2}+y{\cdot}LaAs+y{\cdot}Fe_{2}As
\end{equation}
On the other hand, with presence of entropy, it leads to:
\begin{multline}
\begin{split}
&La_{y}A_{1-y}Fe_{2}As_{2}{\rightarrow} \\
{\frac{1-y}{1-x}}{\cdot}&La_{x}A_{1-x}Fe_{2}As_{2}+{\frac{y-x}{1-x}}{\cdot}LaAs+{\frac{y-x}{1-x}}{\cdot}Fe_{2}As
\end{split}
\end{multline}
where $La_{x}A_{1-x}Fe_{2}As_{2}$ is the solution phase with maximum amount of La dissolved. We define the free energy difference between the two ways of decomposition at temperature $T$ as ${\Delta}G$, which is given by
\begin{equation}
{\Delta}G={\Delta}E-TS
\end{equation}
where ${\Delta}E$ is the energy difference of the two decompositions. In calculating ${\Delta}E$, one needs to calculate the energies of ordered phases (e.g $LaAs$, $Fe_{2}As$), which are estimated by DFT with a supercell model. It also involves computing energy of the solution phase $La_{x}A_{1-x}Fe_{2}As_{2}$, which is estimated with Eq. (1). $S$ is the entropy of the solution phase, which is estimated by Eq. (2)

Evidently, the first decomposition is favored by energy but disfavored by entropy, thus the two decompositions are competing with each other and will eventually reach a balance. At the balance point, we have ${\Delta}G$=0. Then we obtain the following equation to solve the maximum solubility $x$:
\begin{equation}
-2k_{B}T(x{\log}x+(1-x){\log}(1-x))=10x({\Delta}H)
\end{equation}
The factor 10 on the right side results from the fact that there are 10 atoms in each unit cell. ${\Delta}H$ is the formation energy of $LaFe_{2}As_{2}$ or $YFe_{2}As_{2}$ (Table I). Solve this equation at $T$=300 K, we find the solubility $x$ is 20\% for La and $10^{-9}$ for Y. This means, with excess of La, $AFe_{2}As_{2}$ ($A$=Ca, Sr, Ba) will be doped with La up to 20\%. On the other hand, Y is very hard to be doped into $AFe_{2}As_{2}$ at room temperature. Note that the maximum doping concentration only depends on the formation enthalpy of $LaFe_{2}As_{2}$, independent of the solvent, i.e. which cation atoms in $AFe_{2}As_{2}$. Based on analysis above, we argue that for $LaAFe_{4}As_{4}$ ($A$= K, Rb ,Cs), these 1144 structures will maintain the phase until 1000 K or even higher. For $A$=Ca, Sr, Ba, we will get 122(s) phase, with 20\% concentration of La doping. 

\subsection{Electronic Structures}
For the intimate relation between the 1144 phase and the 122 phase, it is interesting to compare their band structures. In previous works, the band structures of various 122 iron arsenides have been studied and several key features have been identified. For example, it has concentric hole-type FS at ${\Gamma}$ point in Brillouin zone (B.Z.) and electronic elliptical FS around M points (The M point in primitive tetragonal B.Z. corresponds to X point in body-centered-tetragonal B.Z. [34]). The density of states near Fermi level is mainly contributed by the Fe 3-$d$ orbitals. The 4$p$ orbitals of As are distributed 2-6 eV below Fermi surface (FS) due to the strong As-Fe interaction [32,33]. Therefore, it is believed that Fe 3$d$-bands are mainly responsible for superconductivity. Previous studies also show that iron-based SC show quasi-two-dimensional FS. [32,33] Thus SC-relevant physics is thought to be two dimensional and can be described by a minimum Hamiltonian model with two bands [40]. 

The band structures of the three most stable $LaAFe_{4}As_{4}$ are calculated and we try to address the following questions. Whether the 2-D nature of FS retains as La is introduced. Whether the band structure keeps the semi-metal feature: hole-pocket at ${\Gamma}$ and electron-pocket at M, which is believed critical for superconductivity. How the cation $A$ in $LaAFe_{4}As_{4}$ will impact the shape of FS. As we know, even with isovalent substitution, the topology of FS might essentially change. For example, the electron-type FS will disappear in $KFe_{2}As_{2}$ [41], in contrast with $RbFe_{2}As_{2}$ [42] and $CsFe_{2}As_{2}$ [43].

\begin{figure}
\includegraphics[scale=0.35]{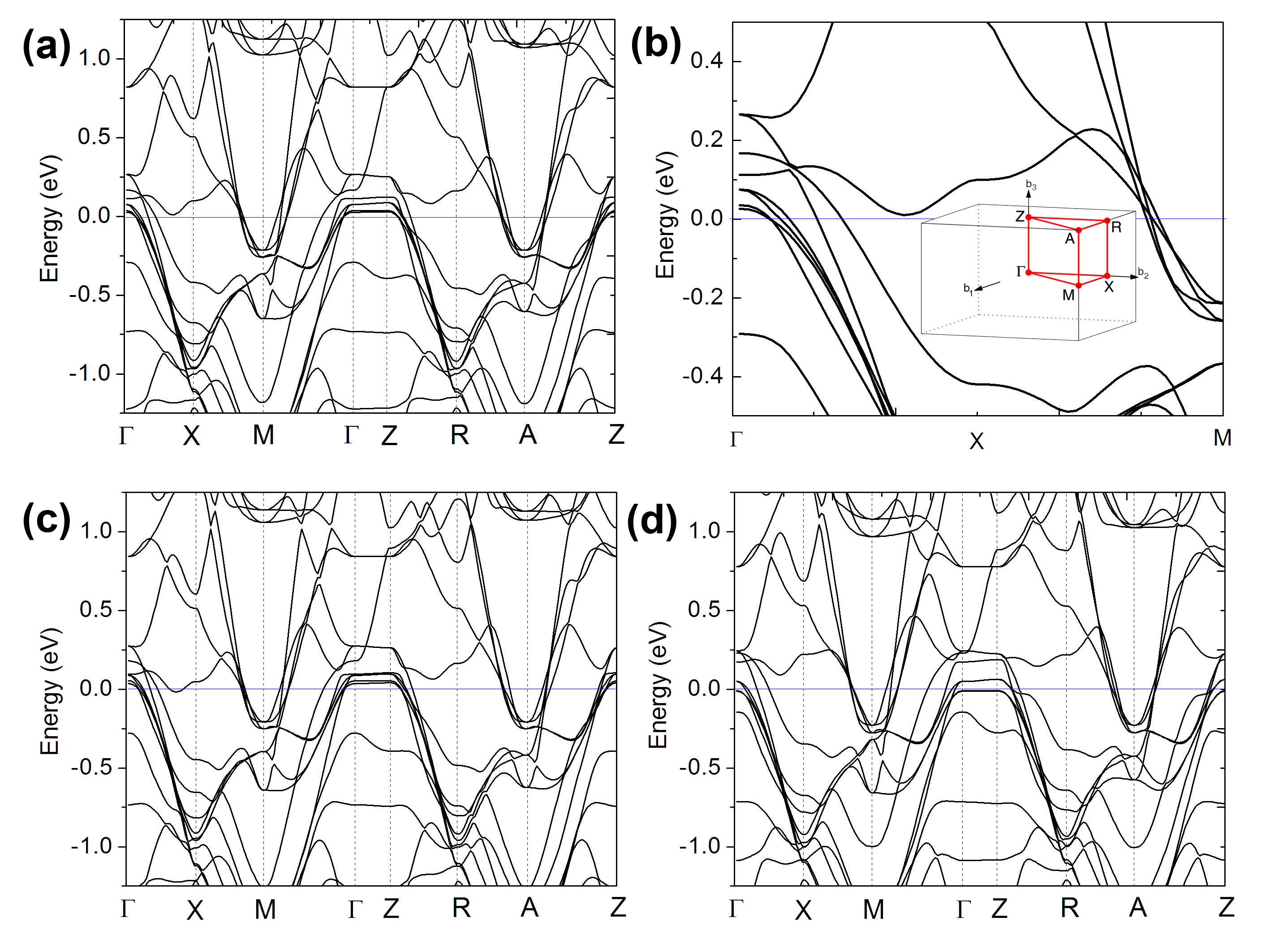}
\caption{\label{fig:epsart} (Color online)(a) Band structure of $LaRbFe_{4}As_{4}$. (b) Zoom-in of (a), in which the insect shows primitive tetragonal B.Z. It clearly shows there are six bands crossing Fermi level at $\Gamma$ point and four bands at M point. (c)(d) The band structures of $LaKFe_{4}As_{4}$ and $LaCsFe_{4}As_{4}$.}
\end{figure}

The band structures of the three 1144 structures are presented in Figure 4, which are showing both similar and distinct features. For example, they all have a hole-type FS at $\Gamma$ point in B.Z. and an electron-type FS at M point. From zoom-in band (Figure 4b), we find there are six bands cross the Fermi level at $\Gamma$ point and four bands cross it at M point. Four of the six bands are closer around the $\Gamma$ point, while the other two bands are further away. We group the six bands into three sets, which are changing differently as varying the $A$ atom in $LaAFe_{4}As_{4}$. The inner four bands are grouped together labelled as “I”. The band in the middle is labelled as “II”, the largest and outer-most band is designated by “III” (Figure 5). In the $k_{z}$ direction, bands I and II have weak dispersion, thus can be considered as 2-D, (Figure 5d, e, f). On the other hand, the FS of band III will “shrink” as going up in $z$ direction and eventually merge into the FS of Band I. In the $k_{x}$-$k_{y}$ plane, the three systems have FS of similar shape, but of different sizes. The FS due to Band I in $k_{z}$=0 plane will shrink as the radius of cation is increased from K to Cs. In particular, two FS sheets in $LaCsFe_{4}As_{4}$ almost shrink to a single point. Band II generally remains unchanged when varying the cations. While FS of Band III will expand as radius of cations increase. On the other hand, the electron-type FS at M point is little affected by the changing of cation $X$. Therefore, substituting cations will mainly give a change to the FS at $\Gamma$ point. 

\begin{figure*}
\includegraphics[scale=0.9]{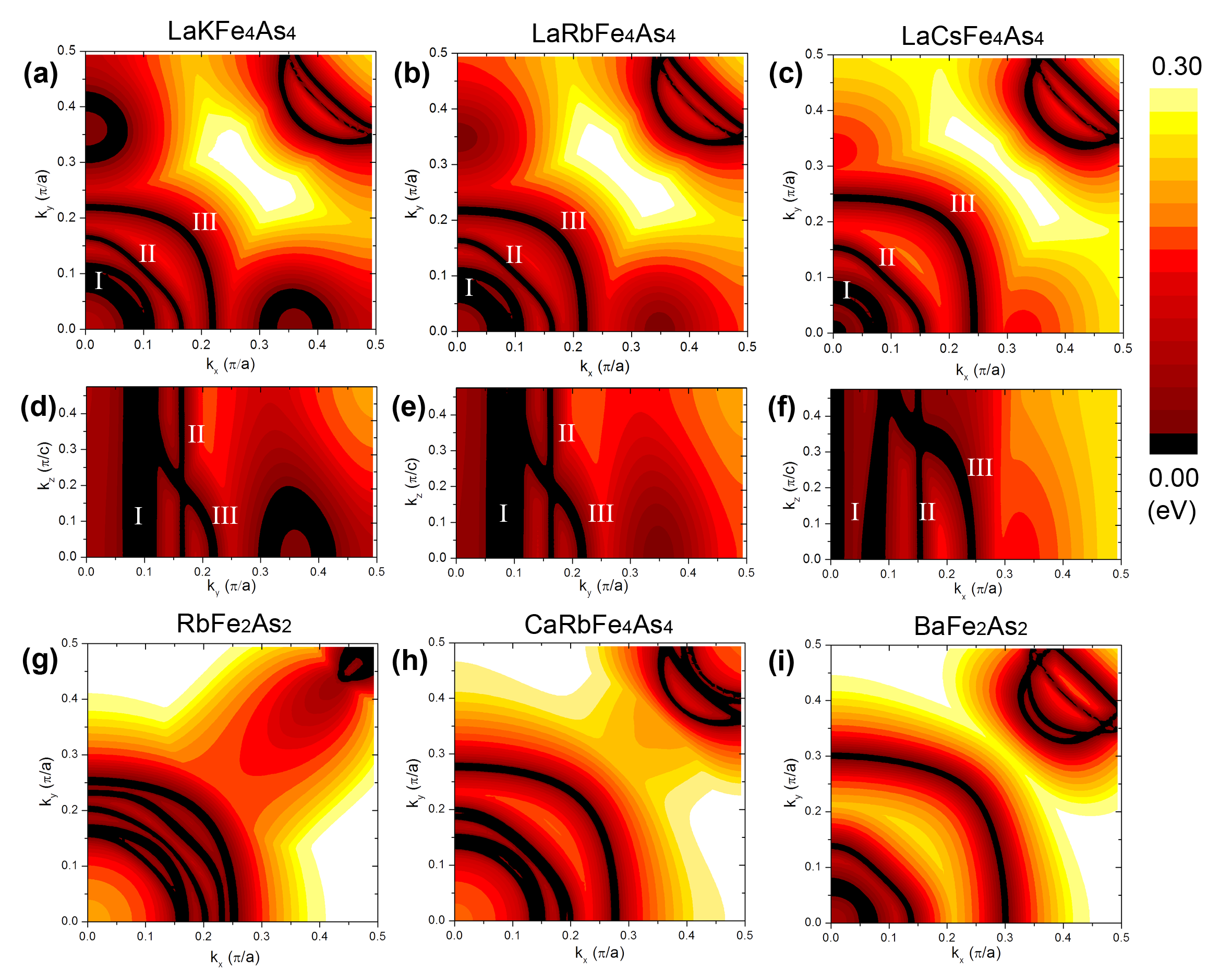}
\caption{\label{fig:epsart} (Color online) (a)-(c) Fermi surface in the $k_{z}$=0 plane for $LaKFe_{4}As_{4}$, $LaRbFe_{4}As_{4}$ and $LaCsFe_{4}As_{4}$. (d)-(f)Fermi surface of the three systems in the $k_{x}$=0 plane. (g)-(i) Fermi surface of $RbFe_{2}As_{2}$, $CaFe_{2}As_{2}$ and $BaFe_{2}As_{2}$ in $k_{z}$=0 plane. Colors in color bar is interpolated with equal intervals in energy.}
\end{figure*}

Special features are found in $LaKFe_{4}As_{4}$: an additional band crossing the FS between the ${\Gamma}$ and X point, resulting in another section of FS. (Figure 4c). The additional FS is of a shape of hollow “bubble” and will disappear along $k_{z}$ direction (Figure 5a,d). On the other hand, in $LaRbFe_{4}As_{4}$ and $LaCsFe_{4}As_{4}$, the band responsible for the “bubble” FS is lying close to but not crossing the FS.

For $LaAFe_{4}As_{4}$ ($A$=Cs Rb K), each unit cell contains two cation atoms and transfer an average of four electrons from cations to the Fe-As layer. In that sense, it might have a similar electronic structure as $BaFe_{2}As_{2}$, which is effectively transferring same amount of charges. In Figure 5i, we present the FS of $BaFe_{2}As_{2}$. We find $BaFe_{2}As_{2}$ having 4 sheets of hole-type FS at $\Gamma$ point and 4 sheets of electron-type FS at M point. While, as mentioned above, $LaRbFe_{4}As_{4}$ have 6 FS sheets at $\Gamma$ and 4 sheets at M. The difference is caused by two bands in $BaFe_{2}As_{2}$ being suppressed in energy and become fully filled, only resulting in four half-filled bands.

$CaRbFe_{4}As_{4}$ has one fewer electron transferred to Fe-As layer. We also compare $LaRbFe_{4}As_{4}$ with $CaRbFe_{4}As_{4}$. Since $LaRbFe_{4}As_{4}$ has one more electron, the Fermi energy should be shifted higher in energy. Our calculation is consistent with expectation, showing a larger FS at M point and smaller FS at $\Gamma$ point. Thus, FS size changes from $CaRbFe_{4}As_{4}$ to $LaRbFe_{4}As_{4}$ can be qualitatively understood by a rigid shift of FS. $CaRbFe_{4}As_{4}$ has 5 bands crossing the FS at $\Gamma$ point and 4 bands crossing FS at M point. Thus providing one more valence electron by La, one is introducing one more section FS at $\Gamma$ point but will not affect the band number at M point. Then we examine $RbFe_{2}As_{2}$, which has two fewer valence electrons compared with $LaRbFe_{4}As_{4}$. $RbFe_{2}As_{2}$ has a larger hole-type FS at $\Gamma$ and small FS at M point, which is qualitatively consistent with the rigid shift of FS. In addition, $RbFe_{2}As_{2}$ have 6 FS sheets at $\Gamma$ and 4 sheets at M point, same as $LaRbFe_{4}As_{4}$.

\section{Main group elements (IIIA)}

In this section, we study In- and Tl-contained 122 or 1144 structures as well as their electronic structures.

\subsection{Phase Stability}
Similar with La and Y, In and Tl do not favor 122 iron arsenides. Our earlier calculation suggests that enthalpy favors a decomposition of $InFe_{2}As_{2}$ into $InAs$ and $Fe_{2}As$. [27] For $TlFe_{2}As_{2}$, it follows a different decomposition: $TlFe_{2}As_{2}{\rightarrow}Tl+2(FeAs)$. Binary alloys of Tl-As are unstable and they will decompose into their elementals (Figure 2b,c). Based on that, the enthalpy difference ${\Delta}H$ is defined as: 
\begin{equation}
\begin{split}
{\Delta}H^{122}&=E_{TlFe2As2}-(E_{Tl}+2{\cdot}E_{FeAs}) \\
{\Delta}H^{122}&=E_{InFe2As2}-(E_{InAs}+E_{Fe2As})
\end{split}
\end{equation}
\begin{table}
\caption{\label{tab:table2} Lattice constants (calculated values) and formation enthalpy of In- and Tl-contained 122 compounds. The last collumn is the valence state of In or Tl in 122 compounds.}
\begin{ruledtabular}
\begin{tabular}{c c c c}
 & Lattice & ${\Delta}H$ & Cation valence \\
 & ($\AA$) & (meV/atom) & \\
$InFe_{2}As_{2}$ & a=3.7242 & 3.12 & +1 \\
				 & c=13.8063 & & \\
$TlFe_{2}As_{2}$ & a=3.7552 & 137.55 & +1 \\
				 & c=13.9856 & & \\
$InCo_{2}As_{2}$ & a=3.7690 & -84.89 & +1 \\
				 & c=13.3537 & & \\
$TlCo_{2}As_{2}$ & a=3.7803 & -267.94 & +1 \\
				 & c=13.6754 & & \\
\end{tabular}
\end{ruledtabular}
\end{table}
The ${\Delta}H$ of $TlFe_{2}As_{2}$ and $InFe_{2}As_{2}$ are listed in Table II. Note that for both Tl and In, the 122 iron arsenide is unstable. This is seemingly another evidence that the +3 cations do not support 122-type iron arsenides. But, orbtial occupancy shows that In or Tl in this environment only lose one of the three valence electrons to the Fe-As layers, leading to +1 valence state (Table II). The higher electronegativity of In and Tl is responsible for such partial loss of valence electrons. We suspect the unstable 122 phase is highly due to the meta-stable +1 cations. Thus, the essential to stabilize the 122 phase is stabilizing the +1 valence state. The solution we propose is injecting extra electrons into Fe-As layers to increase its chemical potential, such that the electron in In orbitals is attracted less by Fe-As and forming +1 valence state will cost less energy. To implement this idea, we substitute Fe with Co. We examine a situation that all Fe are replaced by Co, i.e. $InCo_{2}As_{2}$ and $TlCo_{2}As_{2}$. Based on the phase diagram Figure 2b, the enthalpy difference of 122 $InCo_{2}As_{2}$ is defined as
\begin{equation}
{\Delta}H^{122}=E_{InCo2As2}-(E_{Co}+E_{InAs}+E_{CoAs})
\end{equation}
For Tl, it follows a different way of decomposition: $TlCo2As2{\rightarrow}Tl+2(CoAs)$ (Figure 2c). Thus, the enthalpy of $TlCo_{2}As_{2}$ is defined as
\begin{equation}
{\Delta}H^{122}=E_{TlCo2As2}-(E_{Tl}+2{\cdot}E_{CoAs})
\end{equation}
To estimate the energy, one needs to know the magnetic ordering in the Co-As layer. Cobalt arsenides show complex magnetism. For instance, $BaCo_{2}As_{2}$ and $SrCo_{2}As_{2}$ do not show any magnetic ordering down to 2K. On the other hand, $CaCo_{2}As_{2}$ undergoes a transition to AFM phase at 76K. [37] In addition, these systems might feature complex magnetic fluctuations. For example, $SrCo_{2}As_{2}$ even with a paramagnetic magnetism at T=5K, reveal AFM spin fluctuations, similar as the AFM ordering found in 122 iron arsenide.  

In our DFT study, we have considered six spin configurations (stripe-AFM, FM, etc.) [44] and find that the energies of these configurations are close. This is consistent with paramagnetic ground state as found in experiment. We adopt the following magnetic ordering in our calculation: FM for $BaCo_{2}As_{2}$, and $SrCo_{2}As_{2}$; AFM for $CaCo_{2}As_{2}$ (It means inter-layer AFM, in-plane Co is still FM); for $InCo_{2}As_{2}$, $TlCo_{2}As_{2}$, $KCo_{2}As_{2}$, different spin configurations all converge to a non-magnetic ground state, neither AFM nor FM. The magnetic configuration for the binary alloy has been studied in experiment. For example, $FeAs$ is showing an AFM spiral magnetic ordering, [45]. For $Fe_{2}As$ also show an AFM magnetic ordering. [46] 

The enthalpy of $InCo_{2}As_{2}$, $TlCo_{2}As_{2}$ is listed in Table II. It shows that the stability of 122 phase is remarkably enhanced when Fe is replaced by Co. As far as we know, there has not been much report on the ternary alloy with chemical composition of In-Co-As or Tl-Co-As. Traditional wisdom just suggests there is not ternary alloy (Figure 2b,c). Even our study is not a thorough phase search, nevertheless, it suggests the original phase diagram is incomplete. In addition, $InCo_{2}As_{2}$ and $TlCo_{2}As_{2}$ reveal new ways in constructing 122 structures. The composition of 122 phase has generally followed a “formula”: one non-metal element (e.g P, As, Se, etc.) with electronegativity 2.0-2.5 plus an active metal with an electronegativity $<$ 1.0 (e.g. alkali metals or alkaline earth) and a transition metal with electronegativity $>$ 1.5. Within such chemcial composition, As or P are not powerful enough to seize all the electrons from transition metal and will bond with the transition metal to form the skeleton, and the active metal inserted between the skeleton layer, losing all their valence electrons. Most 122 structures (e.g. $CaFe_{2}As_{2}$, $KFe_{2}As_{2}$) belong to this class. However, in the case of $InCo_{2}As_{2}$ and $TlCo_{2}As_{2}$, we have two metals of high electronegativity: In(Tl) and Co, whose electronegativity values are both greater than 1.5. In that sense, $InCo_{2}As_{2}$ and $TlCo_{2}As_{2}$ are untypical 122 compounds. Above all, our result challenges the traditional wisdom about the In(Tl)-Co-As phase diagram by showing that the 122 ternary phase is more favored in energy than decomposition into binary alloys.

Note that several 122 compounds $XCo_{2}As_{2}$ ($X$=Ca, Sr, Ba, K, Rb, etc.) also exist. It is interesting to ask whether one can obtain 1144 cobalt arsenide by combining two 122. In a similar way as iron 1144, we define the enthalpy for cobalt as:
\begin{equation}
\begin{split}
{\Delta}H^{1144}&=E_{InXCo4As4}-(E_{InCo2As2}+E_{XCo2As2}) \\
{\Delta}H^{1144}&=E_{InXCo4As4}-(E_{TlCo2As2}+E_{XCo2As2})
\end{split}
\end{equation}
The enthalpy of 1144 structures are all listed in Table III. We find two systems might form 1144 structures: $InKCo_{4}As_{4}$ and $InRbCo_{4}As_{4}$. The stability is lower than the 1144 phase found in iron arsenide [27]. In addition, we notice two trends in stability: In or Tl plus IA generally will better stabilize the 1144 structure than IIA elements; smaller X atoms seem to be better to stabilize the 1144 structure.

\begin{figure}
\includegraphics[scale=0.6]{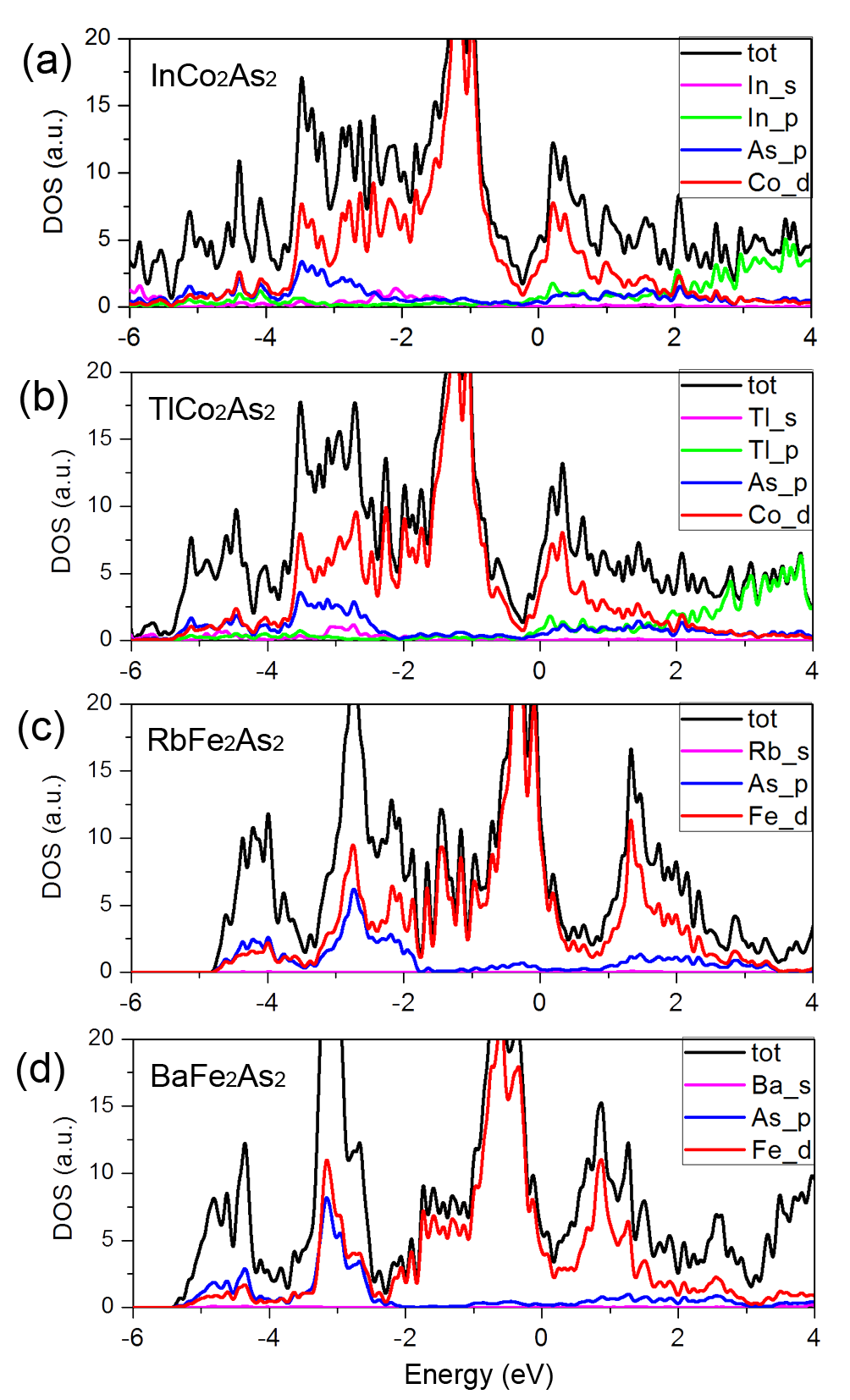}
\caption{\label{fig:epsart} (Color online) The density of states (DOS) and projected-DOS of (a) $InCo_{2}As_{2}$, (b) $TlCo_{2}As_{2}$, (c) $RbFe_{2}As_{2}$ and (d) $BaFe_{2}As_{2}$.}
\end{figure}

\subsection{Electronic Structure}
The density of states (DOS) of $InCo_{2}As_{2}$ and $TlCo_{2}As_{2}$ are presented in Figure 6a, b. The states near the FS is mainly contributed by $d$-orbitals of Co. The $s$-electrons that retains in In(Tl) orbitals lie deep below FS. For In, the peak of $s$-electron is around -2.0 and -6.0 eV; while for Tl the peak is around -3.0 and -4.5 eV. Thus, the occupied $s$-orbital might not directly relate with low-energy excitations. It is also interesting to compare 122 cobalt arsenide with 122 iron arsenide $RbFe_{2}As_{2}$ and $BaFe_{2}As_{2}$ (Figure 6c, d). As cobalt will introduce extra charges, thus FS is expected to be shifted upward. This is exactly what we find in DOS.  The Figure 6a, b clearly show that the FS is located above the valley between the conduction bands and valence bands. While the FS in Figure 6c, d is located below it. Consequently, the density of states for $InCo_{2}As_{2}$ and $TlCo_{2}As_{2}$ are smaller compared with $RbFe_{2}As_{2}$ and $BaFe_{2}As_{2}$. In addition, for $In(Tl)Co_{2}As_{2}$, the $p$-orbital of In(Tl) also make a contribution to the states at FS. On the other hand, in iron arsenide, those states are almost solely contributed by Fe 3$d$-orbtial.  
\begin{figure}
\includegraphics[scale=0.5]{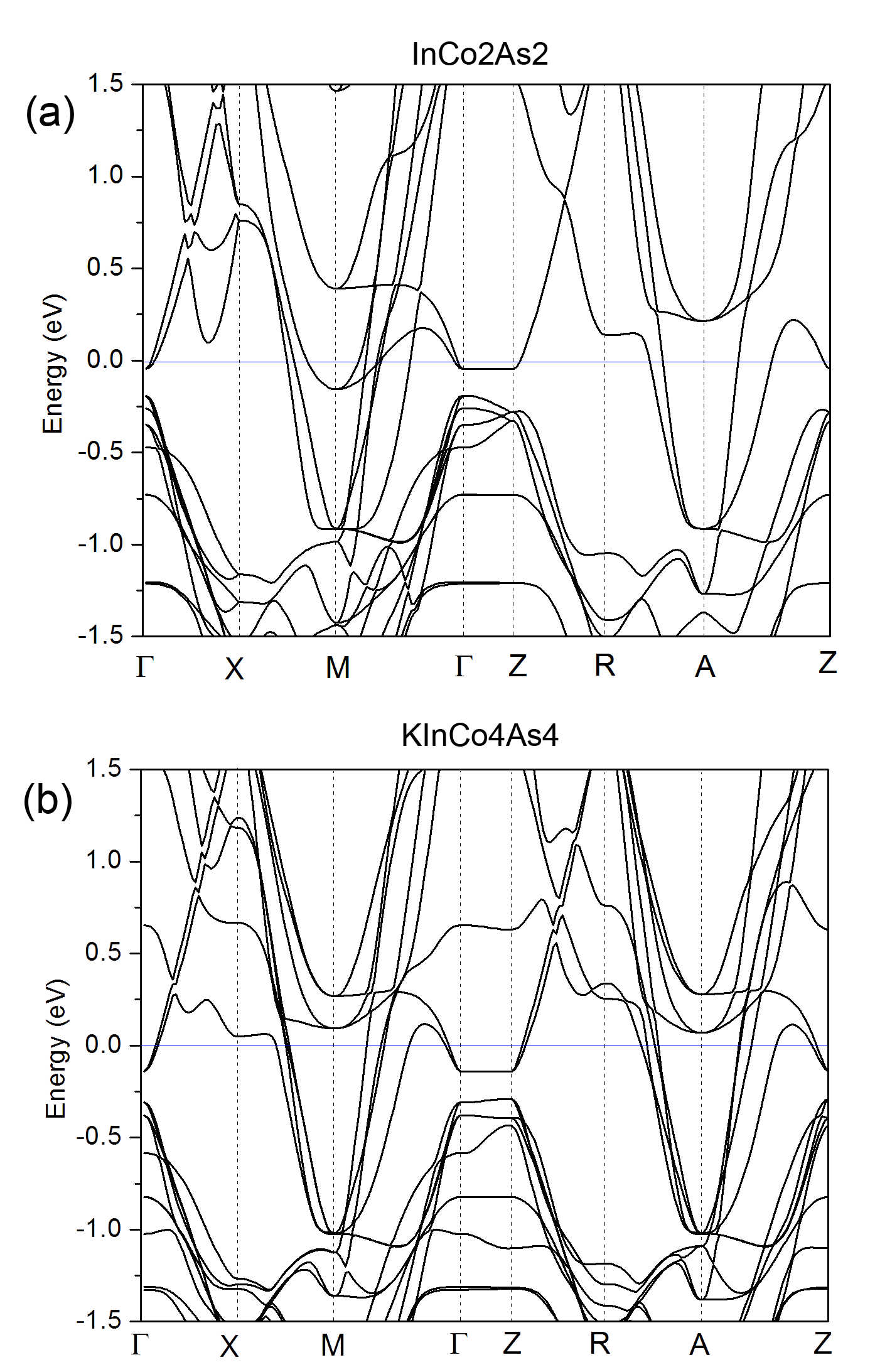}
\caption{\label{fig:epsart} (Color online) The band structure of (a) $InCo_{2}As_{2}$ and (b) $KInCo_{4}As_{4}$.}
\end{figure}

The band structures of $InCo_{2}As_{2}$ are shown in Figure 7a, which features electron-pockets both around the ${\Gamma}$ point and the M point. There are two bands crossing Fermi level at ${\Gamma}$ point and six bands (the graph is showing three double-degeneracy bands) at M point, resulting in multiple FS sheets. The $TlCo_{2}As_{2}$ has almost identical band structure as $InCo_{2}As_{2}$. The FS of the two compounds are plotted in Figure 8. It clearly shows that there is a small FS section around the ${\Gamma}$ point and another set of elliptical FS around the M point. The FS around $\Gamma$ point contains two sheets. Around M point there are six FS sheets, which will form three sets of double degeneracy at the boundary of B.Z. In Figure 8c, d, it is that the FS at $\Gamma$ point showing two-dimensional features and will extend to the top of B.Z. At M point, four of the six FS sheets can be viewed as two-dimensional, as the dispersion in z-direction is weak. On the other hand, the inner FS is bubble-like and will disappear as going up along the $k_{z}$ direction. No surprising that $InCo_{2}As_{2}$ and $TlCo_{2}As_{2}$ are showing similar band structures and FS topology. But Tl has a larger FS at $\Gamma$ point than $InCo_{2}As_{2}$. Note that a relevant 122 compound $SrCo_{2}As_{2}$ shows a couple of complicated FS sheets throughout the B.Z and is not a SC. [35]. Different FS topology might be the reason why $SrCo_{2}As_{2}$ is not a SC. In that sense, $InCo_{2}As_{2}$ is more similar iron arsenide SC and could serve as a test ground for the role of FS played in SC. 
\begin{figure}
\includegraphics[scale=0.6]{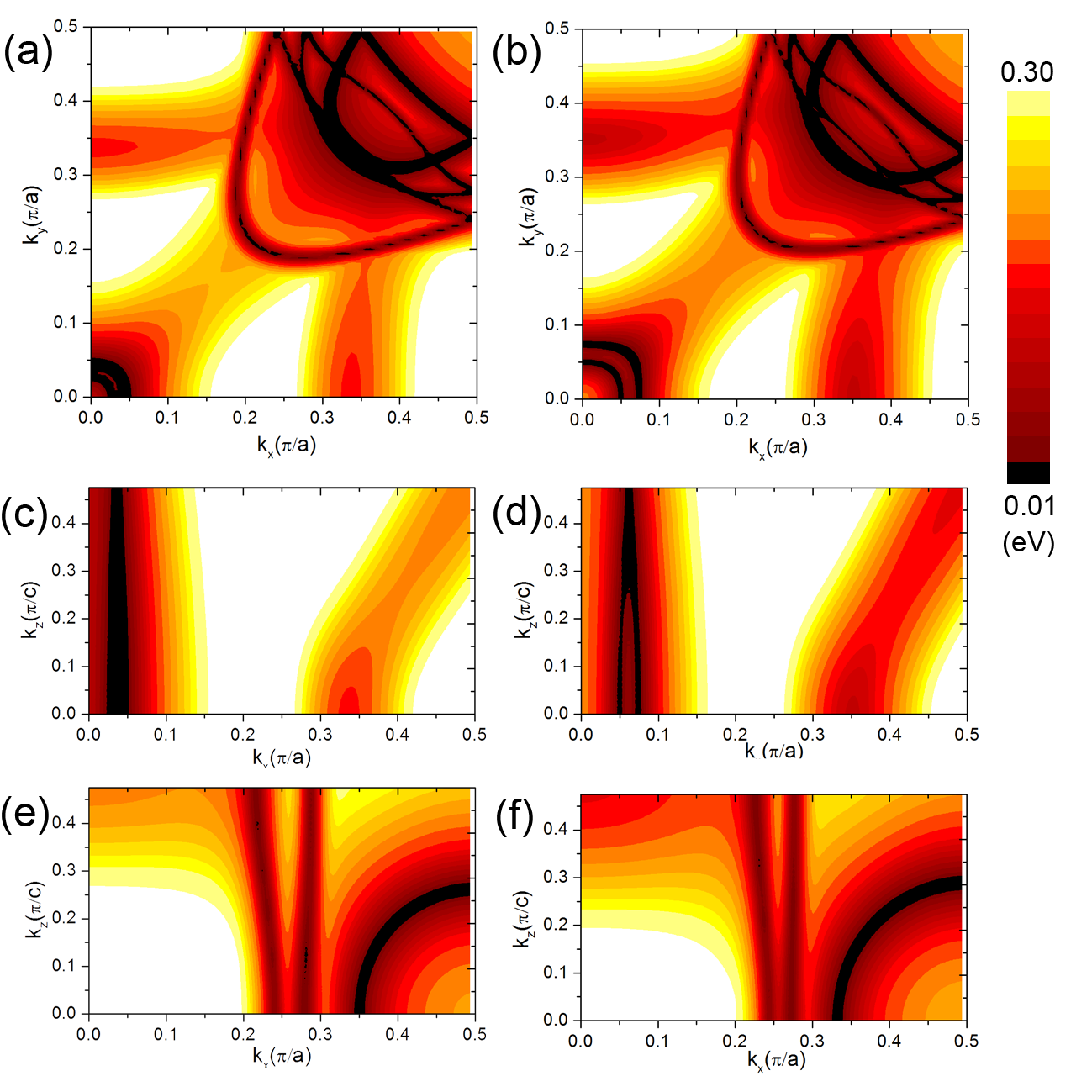}
\caption{\label{fig:epsart} (Color online) The Fermi surface of (a)(c)(e) $InCo_{2}As_{2}$ and (b)(d)(f) $TlCo_{2}As_{2}$ in the $k_{z}$=0, $k_{y}$=0 and $k_{y}$=0.5 plane.}
\end{figure}

The band structure of $InKCo_{4}As_{4}$ is given in Figure7b. $InRbCo_{4}As_{4}$ has very similar band structure as $InKCo_{4}As_{4}$, thus is not shown here. 

\begin{table}
\caption{\label{tab:table3}The lattice constants (calculated values) and enthalpy differences for 1144 structures $InXCo_{4}As_{4}$ and $TlXCo_{4}As_{4}$.}
\begin{ruledtabular}
\begin{tabular}{c c c c c c}
 & Lattice & ${\Delta}H$ & & Lattice & ${\Delta}H$ \\
 & ($\AA$) & (meV/atom) & & ($\AA$) & (meV/atom) \\
InCa & a=3.8223  & 11.76 & TlCa & a=3.8338 & 11.57 \\
	 & c=12.4405 &       &      & c=12.5646 &      \\
InSr & a=3.8148  & 13.17 & TlSr & a=3.8608 & 12.41 \\
	 & c=12.8867 &       &      & c=12.8471 &      \\
InBa & a=3.8377  & 13.11 & TlBa & a=3.8492 & 12.69 \\
	 & c=13.1573 &       &      & c=13.3156 &      \\
InK  & a=3.7690  & -2.28 & TlK  & a=3.7840 & 0.03 \\
	 & c=13.5666 &       &      & c=13.7422 &      \\
InRb & a=3.7879  & -1.05 & TlRb & a=3.7993 & 0.08 \\
	 & c=13.7592 &       &      & c=14.0069 &      \\
\end{tabular}
\end{ruledtabular}
\end{table}

\section{Sm- and Gd-contained 1144 structure.}

Sm and Gd are in neigborhood of Eu in periodic table, similar with Eu in atomic radius and electronegativity. But they do not have stable 122 iron arsenides, as shown in the phase diagram (Figure 2a). Instead, they will decompose: $SmFe_{2}As_{2}{\rightarrow}SmAs+Fe_{2}As$ and $GdFe_{2}As_{2}{\rightarrow}GdAs+Fe_{2}As$. Thus, we define the enthalpy measured with respect to the energy of phase decomposition: 
\begin{equation}
\begin{split}
{\Delta}H^{122}&=E_{SmFe2As2}-(E_{SmAs}+E_{Fe2As}) \\
{\Delta}H^{122}&=E_{GdFe2As2}-(E_{GdAs}+E_{Fe2As})
\end{split}
\end{equation}
Compared with La and Y, Gd and Sm have half-filled $f$-orbital and thus are magnetic in nature. There are several possible magnetic configurations. In our calculation, we have adopted FM configuration as an easy comparison with $EuAFe_{4}As{4}$. We first calculate the enthalpy difference for $SmFe_{2}As_{2}$ and $GdFe_{2}As_{2}$ shown in Table IV. Both $GdFe_{2}As_{2}$ and $SmFe_{2}As_{2}$ are showing positive ${\Delta}H$ (Table IV), which suggests phase decomposition. The unstable 122 phase is a consequence that Eu tends to forms $Eu^{2+}$, but Sm and Gd usually form +3 cations. The enthalpy differences are 66.817 meV/atom for Sm and 83.239 meV/atom for Gd. These values are larger than La but smaller than Y. We also find for both Gd and Sm, they show +3 valence states.

\begin{table}
\caption{\label{tab:table4}Lattice constants and enthalpy difference for 122 structures $SmFe_{2}As_{2}$, $GdFe_{2}As_{2}$ (the first row in table) and 1144 structures $SmAFe_{4}As_{4}$, $GdAFe_{4}As_{4}$.}
\begin{ruledtabular}
\begin{tabular}{c c c c c c}
 & Lattice & ${\Delta}H$ & & Lattice & ${\Delta}H$ \\
 & ($\AA$) & (meV/atom) & & ($\AA$) & (meV/atom) \\
Sm & a=3.7801 & 66.82 & Gd & a=3.7743 & 83.24 \\
	 & c=17.7745 &       &      & c=11.6394 &      \\
SmCa & a=3.8105  & 35.60 & GdCa & a=3.7782 & 49.08 \\
	 & c=11.9532 &       &      & c=11.8991 &      \\
SmSr & a=3.8525  & 31.65 & GdSr & a=3.8218 & 41.71 \\
	 & c=12.1156 &       &      & c=12.2132 &      \\
SmBa & a=3.9103  & 24.90 & GdBa & a=3.8831 & 37.67 \\
	 & c=12.3219 &       &      & c=12.3097 &      \\
SmNa & a=3.7223  & 41.58 & GdNa & a=3.7372 & 42.68 \\
	 & c=12.5030 &       &      & c=12.3742 &      \\
SmK  & a=3.8017  & 13.50 & GdK  & a=3.7903 & 21.51 \\
	 & c=13.0038 &       &      & c=13.0035 &      \\
SmRb & a=3.8589  &  1.77 & GdRb & a=3.8478 &  9.31 \\
	 & c=13.0401 &       &      & c=12.9510 &      \\
SmCs & a=3.8821  & -1.00 & GdCs & a=3.8764 & 4.71 \\
	 & c=13.3146 &       &      & c=13.1842 &      \\
\end{tabular}
\end{ruledtabular}
\end{table}

Since the electronegativity of Sm and Gd is close to La and Y, we use a similar strategy to stabilize 1144 phase. Thus we examine $SmAFe_{4}As_{4}$ and $GdAFe_{4}As_{4}$ ($A$=IA and IIA elements). The enthalpy differences for 1144 phase are defined as:
\begin{equation}
\begin{split}
{\Delta}H^{122}&=E_{SmFe2As2}-(E_{SmAs}+E_{Fe2As}) \\
{\Delta}H^{122}&=E_{GdFe2As2}-(E_{GdAs}+E_{Fe2As})
\end{split}
\end{equation}
We list the enthalpy differences in Table IV. Among various 1144 structures, only $SmCsFe_{4}As_{4}$ showing negative enthalpy. It seems that the 1144 phase can hardly be stabilized. However, we still notice that group IA elements are generally better in stabilizing the 1144 phase than group IIA elements. In addition, as the radius of cation increases, the stability of 1144 improves. This is consistent with the general trends we already found in La- and Y-contained 1144 structures. 

\begin{table}
\caption{\label{tab:table5} Several stable 1144 and 122 structures that contain tri-valence element. The column $\Gamma$ and M show the number of bands crossing Fermi level as well as the FS type (particle or hole). The rightmost column shows the valence state of cations in particular compounds.}
\begin{ruledtabular}
\begin{tabular}{c c c c}
 & $\Gamma$ & M & \\
$LaCsFe_{4}As_{4}$ & 6 (h) & 4 (p) & $La^{3+}$, $Cs^{1+}$ \\
$LaRbFe_{4}As_{4}$ & 6 (h) & 4 (p) & $La^{3+}$, $Rb^{1+}$ \\
$LaKFe_{4}As_{4}$  & 6 (h) & 4 (p) & $La^{3+}$, $K^{1+}$ \\
$RbFe_{2}As_{2}$   & 6 (h) & 4 (p) & $Rb^{+1}$ \\
$CaRbFe_{4}As_{4}$ & 5 (h) & 4 (p) & $Ca^{2+}$, $Rb^{1+}$ \\
$BaFe_{2}As_{2}$   & 4 (h) & 4 (p) & $Ba^{2+}$ \\
$InCo_{2}As_{2}$   & 2 (p) & 6 (p) & $In^{1+}$ \\
$TlCo_{2}As_{2}$   & 2 (p) & 6 (p) & $Tl^{1+}$ \\			 
\end{tabular}
\end{ruledtabular}
\end{table}

\section{Conclusion}

In this work, we examine the possibility of building several tri-valence elements (La, Y, In, Tl, Sm and Gd) into 122/1144 structures. We find that the 1144 phase can be stabilized in three La-contained systems: $LaKFe_{4}As_{4}$, $LaRbFe_{4}As_{4}$ and $LaCsFe_{4}As_{4}$. Two general trends are found: (1) alkali metals elements can better stabilize 1144 $LaXFe_{4}As_{4}$ than alkaline earth elements; (2) larger atom, e.g. Cs will better stabilize 1144 phase than smaller atom K. $LaRbFe_{4}As_{4}$ and $LaCsFe_{4}As_{4}$ show quasi-two-dimensional semi-metal FS, implying superconductivity. Additional bubble-shaped FS is discovered in $LaKFe_{4}As_{4}$. On the other hand, no stable Y-contained 1144 structures are found in our search. For In and Tl, we find $InFe_{2}As_{2}$ and $TlFe_{2}As_{2}$ are not favored in energy, probably because In and Tl are forming the metastable cation states $In^{1+}$ and $Tl^{1+}$ in the environment of the 122 phase. We try to stabilize 122 phase by replacing Fe with Co. Two stable 122 compounds $InCo_{2}As_{2}$ and $TlCo_{2}As_{2}$ are found. Their band structures feature two electron-type pockets around ${\Gamma}$ and M points. For Sm and Gd, most of their 122 or 1144 iron arsenides are unstable. Key information about the band structure of all these compounds are summarized in Table V.

Acknowledgement
We wish to acknowledge the very helpful discussion with Sergey Bud’ko, Tai Kong, William Meier and Mingyu Xu. This work was supported by the U.S. Department of Energy (DOE), Office of Science, Basic Energy Sciences, Materials Science and Engineering Division, including the grant of computer time at the National Energy Research Scientific Computing Center (NERSC) in Berkeley, CA. The research was performed at Ames Laboratory, which is operated for the U.S. DOE by Iowa State University under contract number DE-AC02-07CH11358.   


\begin{thebibliography}{99}
\bibitem{d0det}
Consider three types of in-plane magnetic ordering: FM and stripe-AFM (0, $\pi$), AFM with ($\pi$, $\pi$). Two types of inter-plane magnetic ordering FM and AFM for each of the three in-plane ordering. Thus in all we consider six spin configuraitons.
\end{thebibliography}
\end{document}